\documentclass[preprint,nofootinbib]{revtex4-1}

\usepackage{epsfig}
\usepackage{graphics}\graphicspath{{graph/}{}}
\usepackage{amssymb,amsmath,amsfonts}
\usepackage{time}
\usepackage{color}
\newcommand{\p}{\partial}

\renewcommand{\vec}[1]{\textnormal{\boldmath$#1$}}

\newcommand{\intinf}{\int_{-\infty}^\infty}

\usepackage{hyperref}
\hypersetup{
  colorlinks,
  citecolor=green,
  linkcolor=blue}

\begin{document}
\bibliographystyle{apsrev4-1}

\begin{center}
{\bf\large Long-range correlations in the intrabeam scattering in relativistic beams}


G. Stupakov\\ SLAC National Accelerator Laboratory, 
Menlo Park, CA  94025\\

\vspace{5mm}

\today

\end{center}

%
\section{Introduction\label{sec:1}}
%

The Coulomb scattering in beams of charged particles is an important element of the beam physics in modern accelerators. In plasma, the Coulomb collisions are conventionally described by the collision integral in the Vlasov equation introduced by L. Landau in 1936~\cite{landau36} (see also~\cite{landau_lifshitz_phys_kinetic}). In accelerators, this process is known under the name of the intrabeam scattering (IBS) and has been developed in several papers in the last decades~\cite{Piwinski:1974it, bjorken83m,kubo_oide_IBS,nagaitsev_ibs}.

    \begin{figure}[htb!]
    \centering
    \includegraphics[width=0.5\textwidth, trim=0mm 0mm 0mm 0mm, clip]{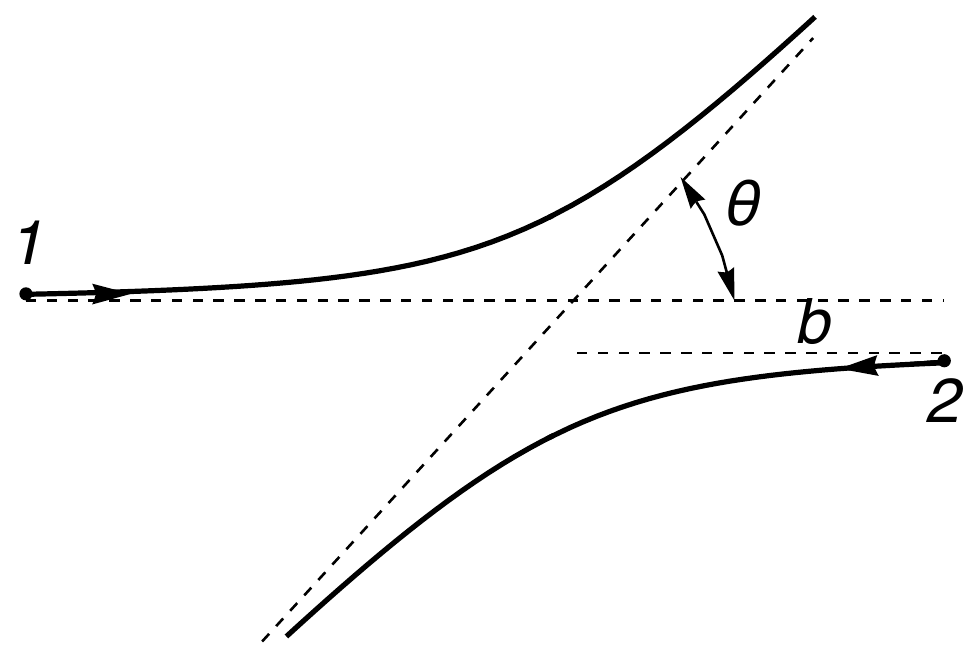}
    \caption{Binary collision of two particles, 1 and 2, in the center-of-mass frame of reference.}
    \label{fig:1}
    \end{figure}
The most straightforward derivation of the collision integral is based on the picture of binary particle collisions in the beam frame, as shown in Fig.~\ref{fig:1}. Before the collision, the two particles are moving along straight orbits and the collision is characterized by the impact parameter $b$. The interaction occurs in a relatively small region where the particles change the direction of motion and after the collision, asymptotically in time, move along  straight orbits rotated by angle $\theta$ relative to the original direction of motion. An important assumption in the derivation of the collision term is that the subsequent binary collisions of the particle are not correlated with each other. This allows one to introduce probabilities and statistical arguments in the analysis and to obtain a relatively simple expression for the collision integral. 

An important simplification in the derivation of the Landau collision integral, and the corresponding expressions for IBS in relativistic beams\footnote{The particle motion in typical accelerator beams in the beam frame of reference is nonrelativistic, so that IBS equations can be obtained from the Landau collision integral by Lorentz transformation from the beam to the laboratory frame.}, comes from the observation that most of the collisions occur at small angles, $\theta \ll 1$. The use of the small-angle approximation limits the accuracy of the theory to the order of $\Lambda^{-1}$ where $\Lambda$ is the Coulomb logarithm,
    \begin{align}\label{eq:1}
    \Lambda
    =
    \ln
    \left(
    \frac{b_\mathrm{max}}{b_\mathrm{min}}
    \right)
    ,
    \end{align}
with $b_\mathrm{max}$ and $b_\mathrm{min}$ the maximum and minimum impact parameters. At the minimum collision parameter $b_\mathrm{min}$ the deflection angle $\theta$ becomes of the order of $\pi /2$ and the small-angle approximation breaks down. The situation with $b_\mathrm{max}$ is more complicated. In plasma, $b_\mathrm{max}$ is traditionally set to the Debye radius, $r_D$, at which the Coulomb interaction force becomes shielded due to the dynamic polarization of charges. This polarization establishes on the timescale that is much larger than the inverse plasma frequency $\omega_p^{-1}$. For relativistic beams, the situation with $b_\mathrm{max}$ is less clear. First, the time $\omega_p^{-1}$ transformed into the laboratory frame could likely correspond to such large distances that the beam motion is perturbed when it passes through the magnets in the lattice. In addition, a formally computed Debye length can be larger than the transverse size of the beam, and then it seems natural that the transverse size of the beam should play the role of $b_\mathrm{max}$. These, and some other arguments in the discussion of the choice of $b_\mathrm{max}$, can be found in Ref.~\cite{kubo_oide_IBS}.

While the picture of binary collisions provides a useful technique for the calculation of collision rates, it falls short of being fully satisfactory. The  problem is that a given particle simultaneously participates in many collisions, making the whole process a collective effect\footnote{In plasma physics, this is manifested by the requirement that the number of particles in the Debye sphere, $N_D = (4\pi /3) r_D^3 n_0$, where $n_0$ is the plasma density, should be large, $N_D \gg 1$, for the Landau collision integral to be valid.}. A framework that takes the nature of the collective collisional interactions into account has been developed many years ago~\cite{Hubbard_1961,hubbard1961p,aono1961}: in it the collisions are treated as the result of the interaction of a particle with the fluctuating electric field created by its neighbors. In this approach, one can obtain a fully converging collision integral without an artificial truncation of the impact parameters by $b_\mathrm{min}$ and $b_\mathrm{max}$ (see, e.g.,~\cite{landau_lifshitz_phys_kinetic}, \S46), however, the result is extremely complicated and  is rearly used in practice.

The goal of this work is to apply the framework of the fluctuating electric field to the IBS taking into account that the transverse motion of the particles in the beam is bounded by the  focusing of the magnetic lattice of the accelerator. This approach naturally incorporates the finite transverse dimensions of the beam and is expected to eliminate the uncertainty associated with the choice of $b_\mathrm{max}$ in conventional theory. We will see however that it also adds new (and quite unexpected!) elements into the description of the Coulomb collisions  which are due to the long-time correlations of the electric field fluctuations in the beam. Such correlations are completely ignored in the  conventional treatment of the IBS.

The paper is organized as follows. In Section~\ref{sec:2}, we formulate a particular IBS problem of energy diffusion that we address in this paper and the simplifying assumptions that we will use in our calculations. In Section~\ref{sec:3}, we express the energy diffusion coefficient $\cal D$ in terms of the integrated correlator of the electric field in the beam. The latter is related to the correlator of the density fluctuations that are calculated in Appendices~\ref{app:A} and~\ref{app:B}. In Section~\ref{sec:4}, we derive a general expression for the averaged over the distribution function coefficient $\bar{\cal D}$ and in Section~\ref{sec:5} we study the limit of this coefficient when the beam energy spread tends to zero. Quite surprisingly, we find that in this limit the integral expression for $\bar{\cal D}$ diverges, although a convergent result can be obtained if one limits the contribution to the integral by small scales of the fluctuating field. The general case of non-zero energy spread is analyzed in Section~\ref{sec:6}. Section~\ref{sec:8} summarizes the results of the paper.

We use the CGS system of units throughout this paper.

%
\section{Formulation of the problem \label{sec:2}}
%

In general, a complete IBS theory involves all three degrees of freedom of the particle. In this paper, however, we will focus only on the energy and will try to calculate the energy diffusion coefficient $\bar {\cal D}_\eta$ defined by the equation
    \begin{align}\label{eq:2}
    \frac{d\sigma_\eta^2}{ds} =  2\bar{\cal D}_\eta
    ,
    \end{align}
where $\sigma_\eta$ is the rms relative energy spread in the beam (in the lab frame), and the bar over $\cal D$ indicates that it is averaged over the beam distribution function. This coefficient can also be used in the diffusion equation for the energy distribution function $f(\eta)$,
    \begin{align}\label{eq:3}
    \frac{\partial f(\eta)}{\partial s}
    =
    \bar{\cal D}_\eta
    \frac{\partial^2 f(\eta)}{\partial \eta^2}
    ,
    \end{align}
where $\eta = \Delta E/E$ is the relative energy deviation. The energy diffusion is important in linear accelerators of freee electon lasers~\cite{huang_ibs02} where it increses the beam energy spread---a critical parameter that defines the gain length in the FEL. Recent experimental studies of the beam energy spread in several FEL facilities emphasized the importance of the correct account of the IBS in the electron injector~\cite{Prat2020,Di_Mitri_2020,Tomin2021,Qian2022,Prat2022} and in some cases indicated a discrepancy between the measured and expected values of the energy spread.

In addition to limiting our analysis to the energy spread only, we will take another important simplifying factor into account: a small ratio of the longitudinal and transverse temperatures, $T_\| /T_\perp$, of the beam in the beam frame. It is straightforward to estimate these temperatures as the transverse and longitudinal (relative to the direction of motion) energy spreads in the beam using the Lorentz transformations:
    \begin{align}\label{eq:4}
    T_\| \sim mc^2 \sigma_\eta^2 
    ,\qquad
    T_\perp \sim \gamma mc^2 \frac{\epsilon_n}{\beta}
    ,
    \end{align}
where  $\beta$ is the beta function of the magnetic lattice, $\epsilon_n$ is the normalized beam emittance, and $\gamma$ is the relativistic Lorentz factor. In many cases, the longitudinal beam temperature turns out to be much smaller than the transverse one, $T_\| \ll T_\perp$. As an example typical for an x-ray FEL, let us take $\sigma_\eta \sim 5\times 10^{-4}$, $\epsilon_n =1\ \mu$m, $\gamma = 2\times 10^3$ and $\beta = 10$ m. This gives $T_\| \approx 0.12$ eV and $T_\perp \approx 100$ eV with $T_\| /T_\perp \sim 10^{-3}$.  To further simplify calculations, we will assume a coasting beam model in which  the beam distribution function does not depend on the longitudinal coordinate $z$. 

In the limit of cold beam, $T_\| \ll T_\perp$, the conventional IBS theory~\cite{lebedev:440, fel11stupakov_2} gives the following expression for the energy diffusion coefficient:
    \begin{align}\label{eq:5} 
    \bar {\cal D}_\eta
    =
    \frac{\sqrt{\pi}}{4}
    \frac{r_e^2}{\gamma^3}
    \frac{\nu}{\epsilon^{3 /2}\beta^{1 /2}}
    \Lambda
    .
    \end{align}
Here $r_e = e^2/ mc^2$ is the classical electron radius, $\nu$ is the linear particle density (the number of particles per unit length), $\epsilon$ is the beam emittance and $\beta$ is the beta function (it is assumed that $\epsilon_x = \epsilon_y = \epsilon$ and $\beta_x = \beta_y = \beta$). Because the energy spread does not enter Eq.~\eqref{eq:5} this equation is also valid when the beam energy spread is equal to zero, $\sigma_\eta = 0$. We will use Eq.~\eqref{eq:5} as a reference for comparison of the result obtained in this work.

%
\section{The kinetic equation, fluctuations and energy diffusion}\label{sec:3}
%

As was pointed out above, we will use the method of fluctuating electric fields to calculate IBS in a relativistic beam. In contrast to the conventional treatment of Ref.~\cite{Hubbard_1961} where the effect of such fields in plasma is computed assuming that particles move along straight orbits, we include in our analysis the curvilinear trajectories of the particles executing betatron oscillations. This naturally takes into account the finite transverse size of the beam and rigorously defines the effective upper limit $b_\mathrm{max}$ in the Coulomb logarithm~\eqref{eq:1}. 

We start from the kinetic equation for the beam in the laboratory frame of reference. The betatron oscillations of the particles are characterized by the action-angle variables $J_x,\phi_x$ and $J_y,\phi_y$ in the horizontal and vertical directions, respectively,
    \begin{align}\label{eq:6}
    x(s)
    =
    \sqrt{2\beta_x(s) J_x}
    \cos[\psi_x(s) + \phi_x]
    ,\qquad
    y(s)
    =
    \sqrt{2\beta_y(s) J_y}
    \cos[\psi_y(s) + \phi_y]
    ,
    \end{align}
where $x(s)$ and $y(s)$ are the horizontal and vertical coordinates of the particle as a function of the longitudinal coordinate $s$ along the beam line, $\beta_x(s)$ and $\beta_y(s)$ are the beta functions in $x$ and $y$ directions, and $\psi_x(s)$ and $\psi_y(s)$ are the corresponding betatron phases. The variables $\phi_x,\phi_y$ are defined in such a way that they, as the actions $J_x,J_y$, are integrals of motion (see, e.g.,~\cite{stupakov2018classical}).

The Vlasov equation is formulated for the 6-dimensional distribution function of the beam, $f(J_x,\phi_x,J_y,\phi_y,z,\eta,s)$, where $s$ is the longitudinal coordinate along the reference orbit, $z = s-v_0t$ is the coordinate that measures the distance inside the beam with $v_0$ the nominal beam velocity, and $\eta = \Delta E /E$ is the relative energy deviation. The distribution function is normalized so that
    \begin{align}\label{eq:7}
    \int_0^\infty dJ_x
    \int_0^{2\pi} d\phi_x
    \int_0^\infty dJ_y
    \int_0^{2\pi} d\phi_y
    \intinf d\eta
    \int_0^L
    dz\,f = 1
    ,
    \end{align}
where $L$  is the bunch length. Because $J_x,\phi_x,J_y,\phi_y$ are integrals of motion, they do not enter the Vlasov equation for $f$ which can be written as
    \begin{align}\label{eq:8}
    \frac{\p f}{\p s}
    +
    \frac{e\delta E_z}{\gamma mc^2}
    \frac{\p f}{\p \eta}
    +
    \frac{\eta}{\gamma^2}
    \frac{\p f}{\p z}
    =
    0
    .
    \end{align}
The electric field $\delta E_z$ in this equation is a fluctuating electric field generated by the random motion of the beam particles; we will calculate its statistical properties in what follows. In our analysis, we ignore the steady state electric field in the beam (the space charge effect) as not relevant to collisions, as well as the fluctuating transverse components of the electric field that would be responsible for the IBS emittance growth effects. The last term in Eq.~\eqref{eq:8} takes into account the longitudinal slippage of the particles caused by the energy deviation $\eta$. This term can be formally eliminated if instead of $z$ we use the variable $\zeta$,
   \begin{align}\label{eq:9}
   z = \zeta + {s\eta}/{\gamma^2},
   \end{align}
and consider $f$ as a function of this variable $\zeta$ instead of $z$, $f(J_x,\phi_x,J_y,\phi_y,\eta,\zeta,s)$. This function satisfies a somewhat simpler equation,
    \begin{align}\label{eq:10}
    \frac{\p f}{\p s}
    +
    \frac{e\delta E_z}{\gamma mc^2}
    \frac{\p f}{\p \eta}
    =
    0
    .
    \end{align}
Of course, all partial derivatives in this equation are taken keeping all other variables constant.

In the next step, we represent $f$ as a sum of the averaged over fluctuations part $f_0$ and a fluctuating (microscopic) part of the distribution function, $\delta f_\mathrm{M}$,
    \begin{align}\label{eq:11}
    f = f_0 + \delta f_\mathrm{M}.
    \end{align}
To guarantee a unique decomposition of $f$ into the two components, we require $\langle \delta f_\mathrm{M} \rangle = 0$, where the angular brackets denote averaging over the fluctuations\footnote{See more details about the separation of $f$ on $f_0$ and $\delta f_\mathrm{M}$ in Appendix~\ref{app:B}.}. The smallness of the fluctuations is  expressed as $|\delta f_\mathrm{M}| \ll f_0$.  Together with $\delta f_\mathrm{M}$, we treat the fluctuating field $\delta E_z$ as a small, first order, quantity, with the averaged value of $\delta E_z$ equal to zero, $\langle \delta E_z \rangle =0$. Substituting Eq.~\eqref{eq:11} into Eq.~\eqref{eq:10}  we obtain
    \begin{align}\label{eq:12}
    \frac{\p f_0}{\p s}
    +    
    \frac{\p \delta f_\mathrm{M}}{\p s}
    +
    \frac{e\delta E_z}{\gamma mc^2}
    \frac{\p f_0}{\p \eta}
    =
    -
    \frac{e\delta E_z}{\gamma mc^2}
    \frac{\p \delta f_\mathrm{M}
    }{\p \eta}
    .
    \end{align}

Averaging Eq.~\eqref{eq:12} over the fluctuations and taking into account that $\langle  \delta f_\mathrm{M}\rangle = \langle  \delta E_z\rangle = 0$, gives the Vlasov equation for the average distribution function $f_0$ with account of collisions,
    \begin{align}\label{eq:13}
    \frac{\p f_0}{\p s}
    =
    \mathrm{St}_\eta
    ,
    \end{align}
where the energy collision integral $\mathrm{St}_\eta$ is,
    \begin{align}\label{eq:14}
    \mathrm{St}_\eta
    =
    -
    \frac{e}{\gamma mc^2}
    \frac{\p}{\p\eta}
    \langle
    \delta E_z
    \delta f_\mathrm{M}
    \rangle
    .
    \end{align}
Subtracting Eq.~\eqref{eq:13} from Eq.~\eqref{eq:12} gives an equation for the first order terms,
    \begin{align}\label{eq:15}
    \frac{\p \delta f_\mathrm{M}}{\p s}
    +
    \frac{e\delta E_z}{\gamma mc^2}
    \frac{\p f_0}{\p \eta}
    =
    0
    ,
    \end{align}
that defines the fluctuating part of the distribution function $\delta f_\mathrm{M}$. 

We will solve Eq.~\eqref{eq:15} by integrating it along the particles' trajectories. The integration is straightforward if we express $\delta E_z(x,y,z,s)$ as a function of the action-angle variables,
    \begin{align}\label{eq:16}
    \delta E_z(x,y,z,s)
    &\to
    \delta E_z(    
    \sqrt{2\beta_x(s) J_x}
    \cos(\psi_x(s) + \phi_x),
    \sqrt{2\beta_y(s) J_y}
    \cos(\psi_y(s) + \phi_y),
    \zeta + s \eta /\gamma^2,s)
    \nonumber\\&
    \equiv
    \delta \tilde E_z( 
    J_x,\phi_x,J_y, \phi_y,\eta,\zeta,s
    )
    ,
    \end{align}
where $\delta \tilde E_z$ is the electric field expressed in terms of the new set of variables. Then from Eq.~\eqref{eq:15} we obtain,
    \begin{align}\label{eq:17}
    \delta f_\mathrm{M}( 
    J_x,\phi_x,J_y, \phi_y,z,s
    )
    =
    -
    \frac{e}{\gamma mc^2}
    \frac{\p f_0}{\p \eta}
    \int_0^s
    ds'\,
    \delta \tilde E_z( 
    J_x,\phi_x,J_y, \phi_y,\eta,\zeta,s'
    )
    .
    \end{align}
Note that in this equation we have ignored the dependence of function $f_0$ versus $s$ and took the derivative $\p f_0 /\p \eta$ outside the integral---this is justified by the fact that the dependence of $f_0$ on $s$ is given by Eq.~\eqref{eq:13} in which the right-hand side, according to Eq.~\eqref{eq:14}, is of the second order of smallness. Substituting Eq.~\eqref{eq:17} into Eq.~\eqref{eq:14} we obtain
    \begin{align}\label{eq:18}
    \mathrm{St}_\eta
    =
    \left(\frac{e}{\gamma mc^2}
    \right)^2
    \frac{\p}{\p\eta}
    \left(
    \frac{\p f_0}{\p \eta}
    \int_0^s
    ds'\,
    \langle    
    \delta\tilde E_z(s)
    \delta \tilde E_z(s')
    \rangle
    \right)
    ,
    \end{align}
(here and below we omit all the arguments in $\delta\tilde E_z$ except for $s$). We expect that when $s\to \infty$ this integral converges to a finite value, which does not depend on $s$, due to the vanishing correlations between $\delta \tilde E_z(s')$ and $\delta \tilde E_z(s)$ when the difference between $s$ and $s'$ becomes large. With this assumption, we can change the integration variable, $s''= s' - s$, extend the lower limit of integration over $s''$ to $-\infty$ and then set $s = 0$:
    \begin{align}\label{eq:19}
    \mathrm{St}_\eta
    =
    \left(\frac{e}{\gamma mc^2}
    \right)^2
    \frac{\p}{\p\eta}
    \left(
    \frac{\p f_0}{\p \eta}
    \int_{-\infty}^0
    ds''\,
    \langle    
    \delta\tilde E_z(0) \delta \tilde E_z(s'')
    \rangle
    \right)
    .
    \end{align}
Comparing this equation with Eq.~\eqref{eq:3}, we obtain an expression for the energy diffusion coefficient,
    \begin{align}\label{eq:20}
    {\cal D}_\eta
    =
    \left(\frac{e}{\gamma mc^2}
    \right)^2   
    \int_{-\infty}^0
    ds\,
    \langle    
    \delta\tilde E_z(0)\delta \tilde E_z(s)
    \rangle 
    .
    \end{align}
Note that this diffusion coefficient ${\cal D}_\eta$ is a function of six variables, $J_x,\phi_x,J_y, \phi_y,\eta,\zeta$, that characterize particle's trajectory in the beam.

%
\section{The energy diffusion coefficient}\label{sec:4}
%

Our next step is to calculate the fluctuating electric field $\delta E_z$ in the beam and to find the correlator $\langle        \delta \tilde E_z(s) \delta\tilde E_z(s') \rangle$. We will do this by expressing $\delta E_z$ through the beam density fluctuations $\delta n$. If the density perturbation is known as a function of coordinates $x,y,z$ and the variable $s$, $\delta n(x, y, z, s)$, it is convenient to work with its Fourier transform $\delta \hat n_{\vec k}(s)$,
    \begin{align}\label{eq:21}
    \delta \hat n_{\vec k}(s)
    =
    \intinf
    dx\,dy\,dz\,
    e^{i\vec k\cdot \vec r}
    \delta n(x,y,z,s)
    ,
    \end{align}
where $\vec k = (k_x ,k_y ,k_z)$. Similar to Eq.~\eqref{eq:21}, we also define the Fourier components of the electric field $\delta \hat E_{z,\vec k}(s)$. The relation between $\delta \hat E_{z,\vec k}$ and $\delta \hat n_{\vec k}$ is derived in Appendix~\ref{app:A} and is given by Eq.~\eqref{eq:A.3}. For the correlator  $\langle \delta \hat E_{z,\vec k}(s) \delta \hat E_{z,\vec k'}(s') \rangle$  this relation gives the following equation:    
    \begin{align}\label{eq:22}
    \langle \delta \hat E_{z,\vec k}(s) \delta \hat E_{z,\vec k'}(s') \rangle
    =
    \frac{(4\pi i)^2 e^2 k_z k'_z}{[\gamma^2(k_x^2+k_y^2)+k_z^2][\gamma^2({k'_x}^2+{k'_y}^2)+{k'_z}^2]}
    \langle \delta \hat n_{\vec k}(s)  \delta \hat n_{\vec k'}(s') \rangle.
    \end{align}
Finally, making the inverse Fourier transform of this expression we can find the field correlator in coordinates $x,y,z,s$
    \begin{align}\label{eq:23}
    \langle
    \delta E_z(x,y,z,s) \delta E_z(x',y',z',s')
    \rangle
    &=
    \frac{1}{(2\pi)^6}
    \intinf d k_x dk_y dk_z d k'_x dk'_y dk'_z    
    \langle \delta \hat E_{z,\vec k}(s) \delta \hat E_{z,\vec k'}(s') \rangle 
    \nonumber\\&\times
    e^{-i (k_x x + k_y y +k_z z + k'_x x' + k_y' y' +k_z' z')}
    .
    \end{align}
To find this correlator as a function of the action-angle variables (which we indicate by the tilde notation),  on the right-hand side of this equation we express the $x,y,z$ through $J_x,\phi_x,J_y,\phi_y,\eta,\zeta$, and $x',y',z'$ through $J_x',\phi_x',J_y',\phi_y',\eta',\zeta'$ variables using Eqs.~\eqref{eq:6} and the relation $z = \zeta + s\eta /\gamma^2$.

An important aspect of fluctuations, as proved in the general theory of collisions in plasma~\cite{landau_lifshitz_phys_kinetic}, is that calculating $\delta n$ and $\delta E_z$ we can ignore the influence of the electric field on particles' orbits and assume that the particles execute free betatron oscillations as they propagate through the magnetic system. The derivation of the correlator $\langle \delta \hat n_{\vec k}  \delta \hat n_{\vec k}' \rangle$ for this case is carried out in Appendix~\ref{app:B} using the formalism of the Klimontovich distribution functions~\cite{klimontovich2013}. In that derivation, we assume that the averaged distribution function $f_0$ of the beam is a Gaussian distribution in energy and exponential distributions in $J_x$ and $J_y$,
    \begin{align}\label{eq:24}
    f_0(J_x,J_y, \eta)
    =
    f_{0x}(J_x)f_{0y}(J_y)f_{0z}(\eta)
    ,
    \end{align}
where
    \begin{align}\label{eq:25}
    f_{0x}(J_x)
    =
    \frac{1}{2\pi\epsilon_{x}}
    e^{-J_x/\epsilon_{x}}
    ,\qquad
    f_{0y}(J_y)
    =
    \frac{1}{2\pi\epsilon_{y}}
    e^{-J_y/\epsilon_{y}}    
    ,
    \qquad  
    f_{0z}(\eta)
    =
    \frac{1}{L\sqrt{2\pi}\sigma_\eta}
    e^{-\eta^2 /2\sigma_\eta^2}
    ,
    \end{align}
with $\epsilon_{x}$ and $\epsilon_{y}$ the emittance in $x$ and $y$ directions, respectively, and $\sigma_\eta$ the rms relative energy spread. Note that $f_0$ does not depend on $\zeta$ and hence represents a (locally) uniform bunch distribution. The auxiliary parameter $L$, representing the system length, does not appear in the final result which is given by Eq.~\eqref{eq:B.8}.

To further simplify the calculations, in what follows, we will assume a smooth focusing channel. This means that the beta functions $\beta_x$ and $\beta_y$ are constant, and the betatron phases are linear functions of the distance, $\psi_x(s) = s /\beta_x$ and $\psi_y(s) = s /\beta_y$. Substituting Eqs.~\eqref{eq:22} and~\eqref{eq:B.8} into Eq.~\eqref{eq:23} then yields,
    \begin{align}\label{eq:26}
    &\langle
    \delta\tilde E_z(s)
    \delta\tilde E_z(s')
    \rangle
    =
    \frac{e^2 \nu}{2\pi^3}
    \int d k_x dk_y d k'_x dk'_y G(k_x,k_y,k'_x,k'_y,s-s')
    \nonumber\\
    &\times
    \exp\left[i(k_x \sqrt{2\beta_x J_x}
    \cos(s /\beta_x + \phi_x) + k_y \sqrt{2\beta_y J_y}
    \cos(s/ \beta_y + \phi_y))\right]
    \nonumber\\
    &\times
    \exp\left[i(k'_x \sqrt{2\beta_x J_x}
    \cos(s' /\beta_x + \phi_x) + k'_y \sqrt{2\beta_y J_y}
    \cos(s' / \beta_y + \phi_y))\right]
    \nonumber\\
    &\times
    \exp
    \left[
    -
    \frac{1}{2}
    \epsilon_x\beta_x
    \left(
    k_x^2
    +
    k_x'^2
    +
    2k_xk_x'\cos[(s'-s) /\beta_x]
    \right)
    \right]
    \nonumber\\
    &\times
    \exp
    \left[
    -
    \frac{1}{2}
    \epsilon_y\beta_y
    \left(
    k_y^2
    +
    k_y'^2
    +
    2k_yk_y'\cos[(s'-s) /\beta_y]
    \right)
    \right]
    ,
    \end{align}
where 
    \begin{align}\label{eq:27}
    G(k_x,k_y,k'_x,k'_y,s-s')
    =
    \intinf  
    dk_z 
    \frac{k_z^2 
    \exp\left[
    -\frac{1}{2}\sigma_\eta^2
    k_z^2{(s-s')^2}/{\gamma^4}
    \right]  
    }{[\gamma^2(k_x^2+k_y^2)+k_z^2][\gamma^2({k'_x}^2+{k'_y}^2)+k_z^2]}
    ,
    \end{align}
and we have carried out the integration over $k_z'$ using the delta function $\delta(k_z+k'_z)$ in Eq.~\eqref{eq:B.8}. Substituting Eq.~\eqref{eq:26} into Eq.~\eqref{eq:20} will give us the diffusion coefficient for a particle trajectory that is characterized by the parameters $J_x,\phi_x,J_y,\phi_y$. While this detailed knowledge might be of use in some particular problems, in many applications it makes sense to average it  using the equilibrium distribution function~\eqref{eq:24}. As a first step in this calculation, we average Eq.~\eqref{eq:26} over the phases $\phi_x, \phi_y$. Given an explicit dependence of the integrand of Eq.~\eqref{eq:26} on $\phi_x$ and $\phi_y$ through the exponential function, this averaging is straightforward. We denote the result of this averaging by double angular brackets:   
    \begin{align}\label{eq:28}
    &\langle\langle
    \delta\tilde E_z(s')
    \delta\tilde E_z(s)
    \rangle\rangle
    =
    \frac{e^2 \nu}{2\pi^3}
    \int d k_x dk_y d k'_x dk'_y G(k_x,k_y,k'_x,k'_y,s-s')
    \nonumber\\
    &\times
    J_0
    \left(
    \sqrt{2\beta_x J_x}
    [k_x^2 + {k'_x}^2 
    +
    2k_x k'_x
    \cos(s -s')/\beta_x]^{{1} /{2}}
    \right)
    \nonumber\\
    &\times
    J_0
    \left(
    \sqrt{2\beta_y J_y}
    [k_y^2 + {k'_y}^2 
    +
    2k_y k'_y
    \cos(s -s')/\beta_y]^{{1} /{2}}
    \right)
    \nonumber\\
    &\times
    \exp
    \left[
    -
    \frac{1}{2}
    \epsilon_x\beta_x
    \left(
    k_x^2
    +
    k_x'^2
    +
    2k_xk_x'\cos[(s'-s) /\beta_x]
    \right)
    \right]
    \nonumber\\
    &\times
    \exp
    \left[
    -
    \frac{1}{2}
    \epsilon_y\beta_y
    \left(
    k_y^2
    +
    k_y'^2
    +
    2k_yk_y'\cos[(s'-s) /\beta_y]
    \right)
    \right]
    ,
    \end{align}
where $J_0$ is the Bessel function of the zero order. In the next step, we average this expression over the action variables integrating it over $J_x$ and $J_y$ with the weight factors $\epsilon_x^{-1} e^{-J/ \epsilon_x}$ and $\epsilon_y^{-1} e^{-J/ \epsilon_y}$. This integration can be carried out using the formula
    \begin{align}\label{eq:29}
    \int_0^\infty
    dx\,
    e^{-x}J_0(a \sqrt{x})
    =
    e^{-a^2 /4}
    ,
    \end{align}
valid for $a>0$. Using the long overline for the result, we obtain
    \begin{align}\label{eq:30}
    &\langle\langle\overline {
    \delta\tilde E_z(s')
    \delta\tilde E_z(s)
    }\rangle\rangle
    =
    \frac{e^2 \nu}{2\pi^3}
    \int d k_x dk_y d k'_x dk'_y G(k_x,k_y,k'_x,k'_y,s-s')
    \nonumber\\
    &\times
    \exp
    \left[
    -
    \epsilon_x\beta_x
    \left(
    k_x^2
    +
    k_x'^2
    +
    2k_xk_x'\cos[(s'-s) /\beta_x]
    \right)
    \right]
    \nonumber\\
    &\times
    \exp
    \left[
    -
    \epsilon_y\beta_y
    \left(
    k_y^2
    +
    k_y'^2
    +
    2k_yk_y'\cos[(s'-s) /\beta_y]
    \right)
    \right]
    .
    \end{align}

Our result can be even more simplified if we make an additional assumption of equal beta functions, $\beta_x = \beta_y = \beta$, and equal emittances, $\epsilon_x = \epsilon_y = \epsilon$. We then have
    \begin{align}\label{eq:31}
    &\langle\langle\overline {
    \delta\tilde E_z(s')
    \delta\tilde E_z(s)
    }\rangle\rangle
    =
    \frac{e^2 \nu}{2\pi^3}
    \int d k_x dk_y d k'_x dk'_y G(k_x,k_y,k'_x,k'_y,s-s')
    \nonumber\\
    &\times    
    \exp
    \left[
    -
    \epsilon\beta
    \left(
    k_x^2
    +
    k_y^2
    +
    k_x'^2
    +
    k_y'^2
    \right)
    \right]
    \exp
    \left[
    -
    2\epsilon\beta
    (k_xk_x' + k_yk_y')\cos[(s'-s) /\beta]
    \right]
    .
    \end{align}
Two more integrations can be carried out if we use the cylindrical coordinate system and write $k_x = k_\perp\cos\alpha, k_y = k_\perp \sin\alpha$ and $k'_x = k'_\perp\cos\beta, k'_y = k'_\perp\sin\beta$,
    \begin{align}\label{eq:32}
    &\langle \langle
    \overline {
    \delta\tilde E_z(s')
    \delta\tilde E_z(s)
    }
    \rangle\rangle
    =
    \frac{e^2 \nu}{2\pi^3}
    \int k_\perp\,dk_\perp\,d\alpha\, k'_\perp\,dk'_\perp\,d\beta\,
    G(k_\perp, k'_\perp,s-s')
    \exp
    \left[
    -
    \epsilon\beta
    \left(
    k_\perp^2
    +
    {k'_\perp}^2
    \right)
    \right]
    \nonumber\\
    &\times
    \exp
    \left[
    -
    2\epsilon\beta
    k_\perp k'_\perp\cos(\alpha - \beta)\cos[(s'-s) /\beta]
    \right]
    ,
    \end{align}
where we have used the fact that $G$ only depends on the absolute values of $k_\perp$ and $k'_\perp$. The integration over $\alpha$ and $\beta$ can be carried out using the formula
    \begin{align}\label{eq:33}
    \int_0^{2\pi}
    d\phi\,
    e^{-a\cos(\phi)}
    =
    2\pi I_0(a)
    .
    \end{align}
The result is,
    \begin{align}\label{eq:34}
    &\langle\langle
    \overline {
    \delta\tilde E_z(s')
    \delta\tilde E_z(s)
    }\rangle\rangle
    =
    \frac{2e^2 \nu}{\pi}
    \int_0^\infty 
    dk_\perp\,dk'_\perp\,
    G(k_\perp, k'_\perp, s-s')
    \nonumber\\
    &\times
    {k_\perp k'_\perp}
    \exp
    \left[
    -
    \epsilon\beta
    \left(
    k_\perp^2
    +
    {k'_\perp}^2
    \right)
    \right]
    I_0(
    2\epsilon\beta
    k_\perp k'_\perp \cos[(s'-s) /\beta]
    )
    .
    \end{align}
We now introduce the dimensionless variables $\kappa = \sqrt{\epsilon \beta}\, k_\perp$ and $\xi = s /\beta$, which cast Eq.~\eqref{eq:34} into the following one
    \begin{align}\label{eq:35}
    \langle\langle\overline {
    \delta\tilde E_z(s')
    \delta\tilde E_z(s)
    }\rangle\rangle
    &=    
    \frac{2\nu e^2}{\pi\gamma (\epsilon \beta)^{3/ 2}}
    \int_0^\infty 
    d\kappa\,d\kappa'\,G(\kappa,\kappa',\xi-\xi')
    \nonumber\\&\times
    {\kappa\kappa'}
    \exp
    \left[
    -
    \left(
    \kappa^2
    +
    \kappa'^2
    \right)
    \right]
    I_0(
    2
    \kappa\kappa'\cos(\xi'-\xi)
    )
    ,
    \end{align}
with
    \begin{align}\label{eq:36}
    G(\kappa,\kappa',\xi)
    =
    \intinf  
    dq_z 
    \frac{q_z^2 
    \exp\left(
    -\frac{1}{2}\sigma_\eta^2
    q_z^2
    \xi^2 \beta/\epsilon\gamma^2
    \right)
    }{(\kappa^2+q_z^2)(\kappa'^2+q_z^2)}
    ,
    \end{align}
where $q_z = \sqrt{\epsilon \beta} k_z /\gamma$. 

Recalling Eq.~\eqref{eq:20}, we now have the averaged diffusion coefficient $\bar{{\cal D}}_\eta$,
    \begin{align}\label{eq:37}
    \bar{{\cal D}}_\eta
    &=      
    \frac{2\nu r_e^2}{\pi\gamma^3 \epsilon^{3/ 2} \beta^{1/ 2}}
    \int_0^\infty 
    d\kappa\,d\kappa'
    \int_0^{\infty}
    d\xi\,
    G(\kappa,\kappa',\xi)
    \nonumber\\&\times
    {\kappa\kappa'}
    \exp
    \left[
    -
    \left(
    \kappa^2
    +
    \kappa'^2
    \right)
    \right]
    I_0(
    2
    \kappa\kappa'\cos(\xi)
    )
    .
    \end{align}

%
\section{Cold beam limit and small-scale fluctuations}\label{sec:5}
%

As we mentioned in the Introduction, Eq.~\eqref{eq:5} gives the averaged energy diffusion coefficient for a cold beam derived from the conventional IBS theory. It is natural to try to reproduce that formula in our formalism setting $\sigma_\eta = 0$ in Eq.~\eqref{eq:37}. In this limit, the function $G$~\eqref{eq:36} simplifies,
    \begin{align}\label{eq:38}
    G(\kappa,\kappa',\xi')
    =
    \intinf  
    dq_z 
    \frac{q_z^2 
    }{(\kappa^2+q_z^2)(\kappa'^2+q_z^2)}
    =
    \frac{\pi}{\kappa + \kappa'}    
    ,
    \end{align}
and gives for the diffusion coefficient~\eqref{eq:37} the following expression:
    \begin{align}\label{eq:39}
    \bar {\cal D}_\eta
    &=
    \frac{2\nu r_e^2}{\pi\gamma^3 \epsilon^{3/ 2} \beta^{1/ 2}}
    \int_0^{\infty}
    d\xi\,
    \int_0^\infty 
    d\kappa\,d\kappa'    
    \frac{\kappa\kappa'}{\kappa+\kappa'}
    \exp
    \left[
    -
    \left(
    \kappa^2
    +
    \kappa'^2
    \right)
    \right]
    I_0(    2    \kappa\kappa'\cos(\xi)    )      
    .
    \end{align}
Unfortunately, the integral on the right-hand side of the equation diverges at $\xi \to \infty$ because $I_0( 2 \kappa \kappa' \cos(\xi))$ is a periodic function of the variable $\xi$ and does not vanish at infinity. We will return to the discussion of physical mechanism of this divergence in the next section.

Nevertheless, it turns out that we can obtain the diffusion coefficient~\eqref{eq:5} from Eq.~\eqref{eq:39} if we limit the integration in Eq.~\eqref{eq:39} by the wavenumbers that are much larger than the inverse bunch size $(\epsilon \beta)^{-1/2}$. In other words, we will only take into account small scale fluctuations of the electric field.  In terms of  our dimensionless variables $\kappa$ and $\kappa'$, it means that we limit the integration by the region $\kappa, \kappa' \gg 1$. In this limit, we can use the asymptotic expression of $I_0$ for a large argument,
    \begin{align}\label{eq:40}
    I_0(x) \approx ({2\pi |x|})^{-1/2}e^{|x|}.
    \end{align}
Substituting this expression into the internal integral in Eq.~\eqref{eq:39} we obtain,
    \begin{align}\label{eq:41}
    \int 
    d\kappa\,d\kappa'\,
    \frac{\kappa\kappa'}{\kappa+\kappa'}
    \exp
    \left[
    -
    \left(
    \kappa^2
    +
    \kappa'^2
    -
    2\kappa\kappa'|\cos(\xi)|
    \right)
    \right]
    \frac{1}{\sqrt{4\pi \kappa\kappa'|\cos(\xi)|}}
    .
    \end{align}
In this integral, we do not specify the integration limits and will return to their choice below. The exponential function in the integrand is exponentially small for most values of $\kappa \gg 1$ and $\kappa' \gg 1$, unless $\kappa^2 + \kappa'^2$ is almost cancelled by $2\kappa\kappa'|\cos(\xi)|$. This happens when $\kappa'$ is close to $\kappa$: $\kappa' = \kappa + q$ with $|q| \ll \kappa$, and $\xi$ is close to $\pi n$, where $n$ is an integer (for $\kappa' = \kappa$ and $|\cos(\xi)| = 1$, the argument of the exponential function is equal to zero). Each such occurrence can be considered as an analog of a two-particle collision in the framework of the binary collisions. Let us consider one value of $n$, and write $\xi = \pi n + \delta \xi$ with $|\delta \xi| \ll 1$. We then have
    \begin{align}\label{eq:42}
    |\cos(\xi)|
    =
    |\cos(\pi n + \delta \xi)|
    \approx
    1 - \frac{1}{2}\delta \xi^2
    ,
    \end{align}
and the integral Eq.~\eqref{eq:41} is approximated as
    \begin{align}\label{eq:43}
    &\frac{1}{4\sqrt{\pi}}
    \int 
    d\kappa\, d\kappa'\,
    \exp
    \left[
    -
    \left(
    \kappa^2
    +
    \kappa'^2
    \right)
    +
    2\kappa\kappa'\left(1 - \frac{1}{2}\delta \xi^2\right)
    \right]
    \nonumber\\
    &=
    \frac{1}{4\sqrt{\pi}}
    \int d\kappa
    \int_{-\infty}^\infty 
    dq\,
    \exp
    \left[
    -
    q^2
    -
    \kappa^2\delta \xi^2
    \right]
    =
    \frac{1}{4}
    \int    d\kappa
    \exp
    \left(
    -
    \kappa^2\delta \xi^2
    \right)
    .
    \end{align}
Note that since we assume $\kappa \gg 1$, the integral over $q$ in this equation converges over the scale $q \sim 1$ which is much smaller than $\kappa$, and for this reason we extended the integration over $q$ from $-\infty$ to $+\infty$.  

We now need to substitute  this expression to the integral in Eq.~\eqref{eq:39}. To compare the result with the standard collision term, we consider only the ``first collision'', that is $n=0$ in Eq.~\eqref{eq:42}. The integration over $\xi$ in Eq.~\eqref{eq:39} is replaced by the integration over $\delta \xi$ and because the integral converges rapidly when $\delta \xi \to \infty$, the integration interval can be extended to infinity:
    \begin{align}\label{eq:44}
    \bar {\cal D}_\eta
    &=
    \frac{\nu r_e^2}{2\pi\gamma^3 \epsilon^{3/ 2} \beta^{1/ 2}}
    \int
    d\kappa
    \int_0^\infty
    d\delta \xi\,
    \exp
    \left(
    -
    \kappa^2\delta \xi^2
    \right)
    \nonumber\\
    &=
    \frac{\sqrt{\pi}}{4}
    \frac{\nu r_e^2}{\gamma^3 \epsilon^{3/ 2} \beta^{1/ 2}}
    \int
    \frac{d\kappa}{\kappa}
    .
    \end{align}
This result coincides with Eq.~\eqref{eq:5} if we limit the integration over $\kappa$ between some $\kappa_\mathrm{min}$ and $\kappa_\mathrm{max}$ and associate the Coulomb logarithms with such integral:
    \begin{align}\label{eq:45}
    \Lambda
    =
    \int_{\kappa_\mathrm{min}}^{\kappa_\mathrm{max}}
    \frac{d\kappa}{\kappa}
    .
    \end{align}
Since we assumed in our derivation $\kappa \gg 1$, $\kappa_\mathrm{min}$ should also be large, $\kappa_\mathrm{min} \gg 1$.  This means that we take into account the fluctuations in the beam with characteristic scales much smaller than the beam size. Comparing Eq.~\eqref{eq:45} with Eq.~\eqref{eq:1}, we see that $\kappa_\mathrm{min}$ can be associated with the inverse maximum impact parameter $b_\mathrm{max}$ (normalized by the beam size), $\kappa_\mathrm{min} \sim \sqrt{\epsilon \beta}/b_\mathrm{max}$. This also tells us that 
$\kappa_\mathrm{max} \sim \sqrt{\epsilon \beta}/b_\mathrm{min}$. The fluctuations on this scale correspond to the large-angle collisions and cannot be properly treated by our theory that assumes a small $\delta E_z$ and uses a perturbation theory based on this smallness.

%
\section{Calculation of the diffusion coefficient for $\sigma_\eta \ne 0$}\label{sec:6}
%

We now drop the assumption $\sigma_\eta = 0$ and consider the general case given by Eq.~\eqref{eq:37} which we write as
    \begin{align}\label{eq:46}
    \bar {\cal D}_\eta
    &=
    \frac{2\nu r_e^2}{\pi\gamma^3 \epsilon^{3/ 2} \beta^{1/ 2}}
    J    
    ,
    \end{align}
where
    \begin{align}\label{eq:47}
    J&
    =
    \int_0^\infty 
    d\kappa\,d\kappa'
    \int_0^{\infty}
    d\xi\,
    G(\kappa,\kappa',\xi)
    {\kappa\kappa'}
    \exp
    \left[
    -
    \left(
    \kappa^2
    +
    \kappa'^2
    \right)
    \right]
    I_0(
    2
    \kappa\kappa'\cos(\xi)
    )
    \\\nonumber
    &=
    \int_0^\infty 
    d\kappa\,d\kappa'
    \int_0^{\infty}
    d\xi\,
    \intinf  
    dq_z 
    \frac{q_z^2 
    \exp\left(
    -    
    \tau q_z^2 \xi^2
    \right)
    }{(\kappa^2+q_z^2)(\kappa'^2+q_z^2)}    
    {\kappa\kappa'}
    \exp
    \left[
    -
    \left(
    \kappa^2
    +
    \kappa'^2
    \right)
    \right]
    I_0(
    2
    \kappa\kappa'\cos(\xi)
    )    
    ,
    \end{align}
with $\tau = \frac{1}{2}\sigma_\eta^2 \beta/\epsilon\gamma^2$. Even though we set the upper limits of the integration over $\kappa$ and $\kappa'$ to infinity,  we will later need to limit it by some finite value $\kappa_\mathrm{max}$, as we did in Eq.~\eqref{eq:44}. For numerical calculations of the integral in Eq.~\eqref{eq:47}, we will use the cylindrical coordinates in $(\kappa, \kappa')$ space: $\kappa =  K \cos{\alpha}$, $\kappa' =  K \sin{\alpha}$, $d\kappa\,d\kappa' =  K dK\, d\alpha$, and represent the four-dimensional integral $J$ as a combination of one and two-dimensional ones:
    \begin{align}\label{eq:48}
    J
    &=
    \int_0^\infty 
    d K
    P(K)  
    ,
    \end{align}
where
	\begin{align}\label{eq:49}
	    P(K)
    &=
    \int_0^{\infty}
    d\xi\,
    Q(\xi, K) 
    ,
	\end{align}
and
    \begin{align}\label{eq:50}
    Q(\xi, K)  
    &=
    K^3
    \int_0^{\pi /2}
    d\alpha
    \intinf  
    dq_z 
    \frac{q_z^2 
    \exp\left(
    -    
    \tau q_z^2 \xi^2
    \right)
    }{( K^2 \cos^2{\alpha}+q_z^2)( K^2 \sin^2{\alpha}+q_z^2)}    
    \nonumber\\&\times     
    \cos{(\alpha)}\sin{(\alpha)}\,
    e^{-
     K^2}
    I_0[
    2
     K^2 \cos{(\alpha)}\sin{(\alpha)}\cos(\xi)
    ]  
    .
    \end{align}
The function $Q$ can be interpreted as the correlation of fields with the wavenumber $\sim K$ at the distance $s = \beta \xi$.

Unfortunately, not much can be done analytically with the integral Eq.~\eqref{eq:48}. Of course, for a given set of beam and lattice parameters, the diffusion coefficient can be calculated numerically. To illustrate such a calculation, we consider an electron beam with the energy corresponding to $\gamma = 2000$, the normalized transverse emittance $\epsilon_n = 1\ \mu$m and the relative energy spread $\sigma_\eta = 5\times 10^{-4}$. The beam propagates in a focusing channel with $\beta = 8$ m. For the dimensionless parameter $\tau$, this gives $\tau = 5\times 10^{-4}$.

\begin{figure}[htb!]
    \centering
    \includegraphics[width=0.5\textwidth, trim=0mm 0mm 0mm 0mm, clip]{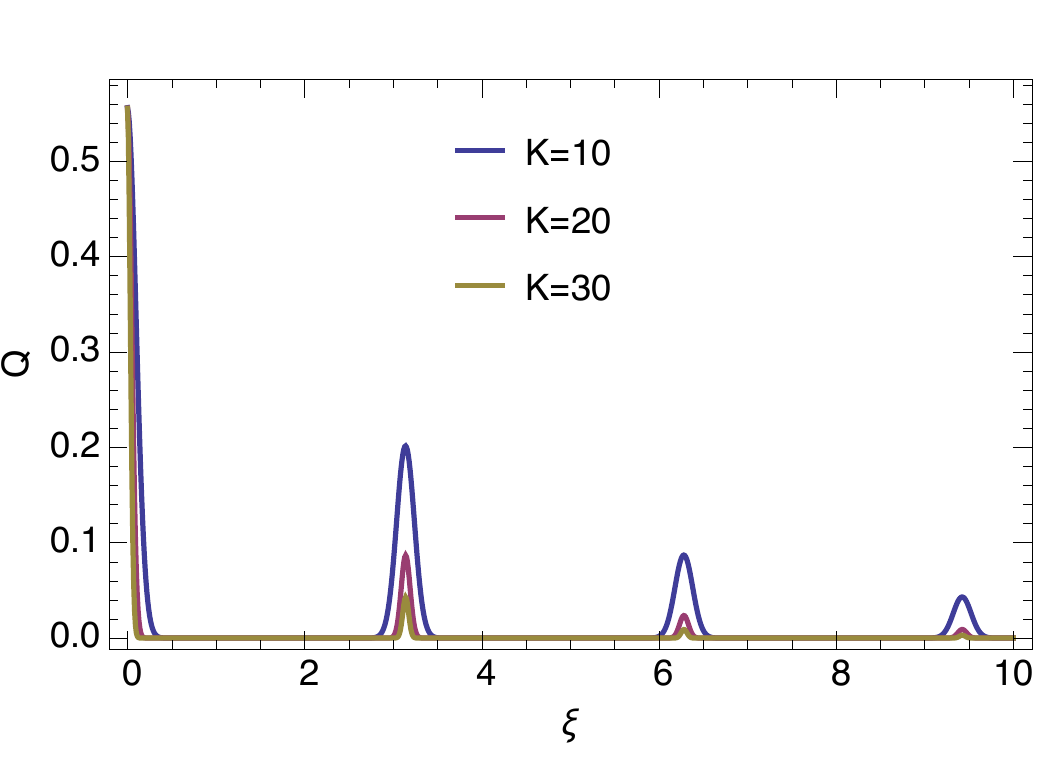}
    \caption{The plot of the function $Q$ versus $\xi = s/\beta$ for a beam with $\tau = 5\times 10^{-4}$ for three different values of $K$.}
    \label{fig:2}
    \end{figure}

\begin{figure}[htb!]
    \centering
    \includegraphics[width=0.5\textwidth, trim=0mm 0mm 0mm 0mm, clip]{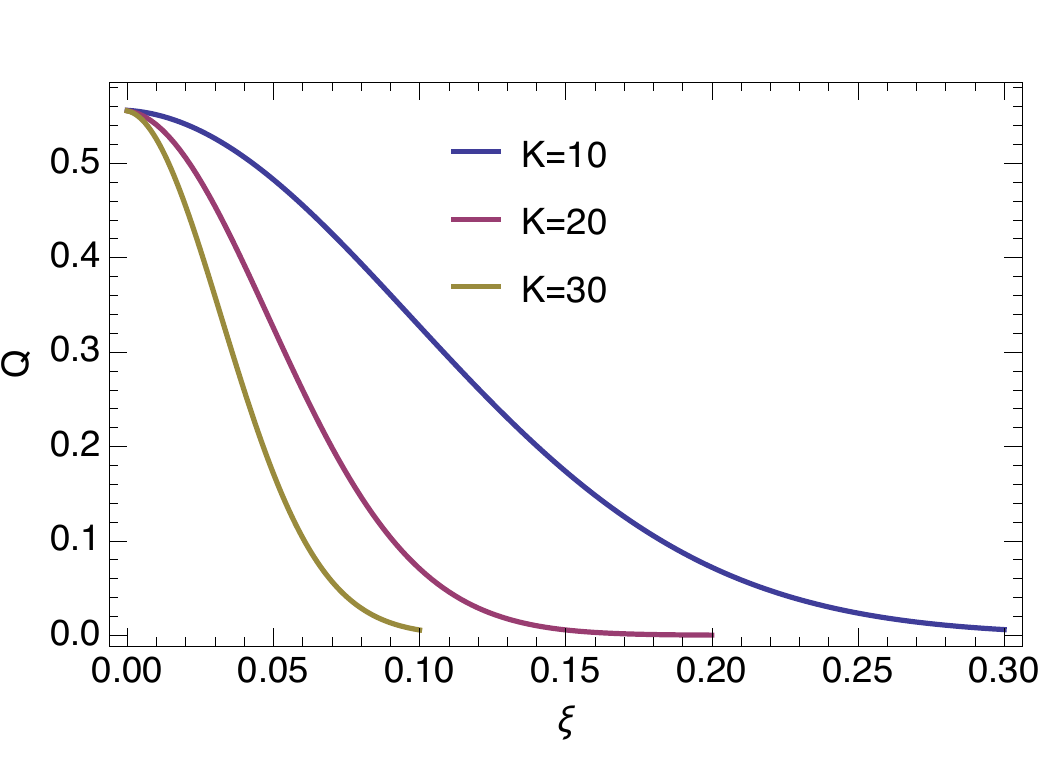}
    \caption{The function $Q$ at small values of $\xi = s/\beta$.}
    \label{fig:3}
    \end{figure}

First, we calculate the function $Q(\xi, K) $ using the two-dimensional integral Eq.~\eqref{eq:50}. The plot of this function for three different values of $K = 10, 20, 30$ is shown in Fig.~\ref{fig:2}. We see that the field correlations quickly decrease almost to zero near the origin, at $\xi \ll 1$,  but then reappear as sharp spikes at $\xi \approx \pi, 2\pi, 3\pi,\ldots$. The amplitudes of the subsequent spikes become smaller, with the spikes corresponding to larger values of $K$ decaying faster than those with smaller values of $K$. The plot of $Q$ in the region of the first spike, at $\xi < 0.3$ is shown in Fig.~\ref{fig:3}.

\begin{figure}[htb!]
    \centering
    \includegraphics[width=0.5\textwidth, trim=0mm 0mm 0mm 0mm, clip]{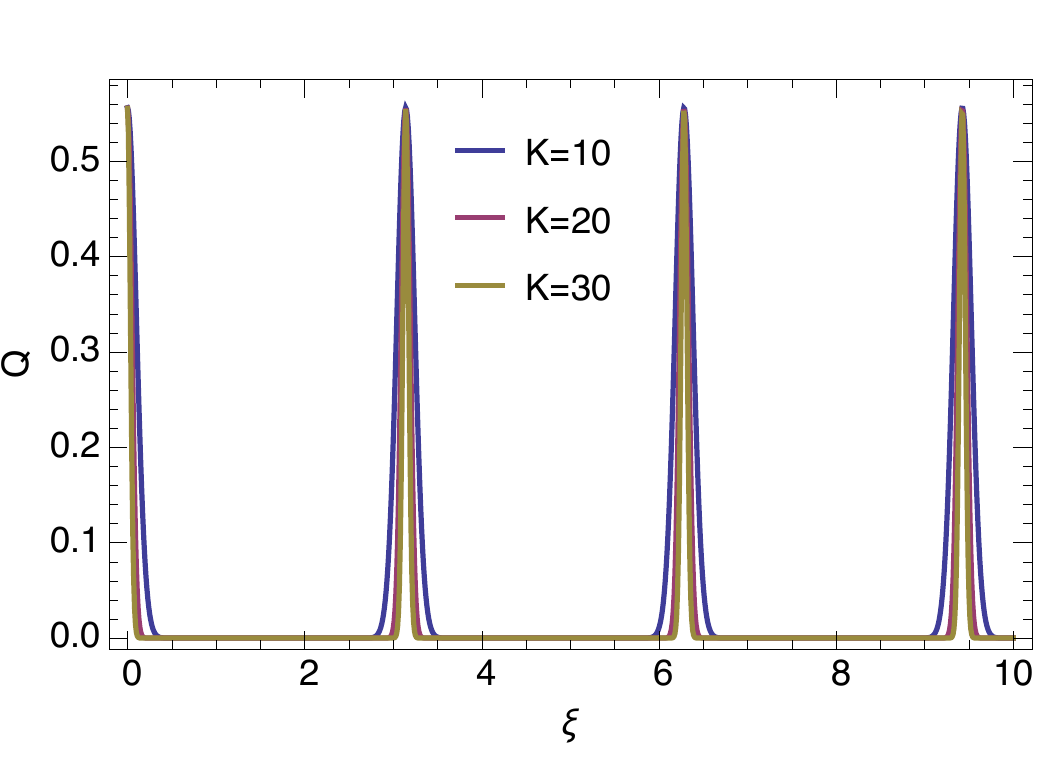}
    \caption{The plot of the function $Q$ versus $\xi = s/\beta$ for a cold beam with $\tau = 0$ for three different values of $K$.}
    \label{fig:4}
    \end{figure}

For comparison, in Fig.~\ref{fig:4} we show the function $Q$ for the same beam parameters, except for the value of $\sigma_\eta$ set to zero, $\sigma_\eta = 0$, that is in the limit of the cold beam.    
As we see, in this case the amplitude of the subsequent spikes remains constant. Since the function $Q$ needs to be integrated over $\xi$ from zero to infinity, it is clear that such an integral would diverge. This is the situation that we have discussed in Section~\ref{sec:5}.

\begin{figure}[htb!]
    \centering
    \includegraphics[width=0.5\textwidth, trim=0mm 0mm 0mm 0mm, clip]{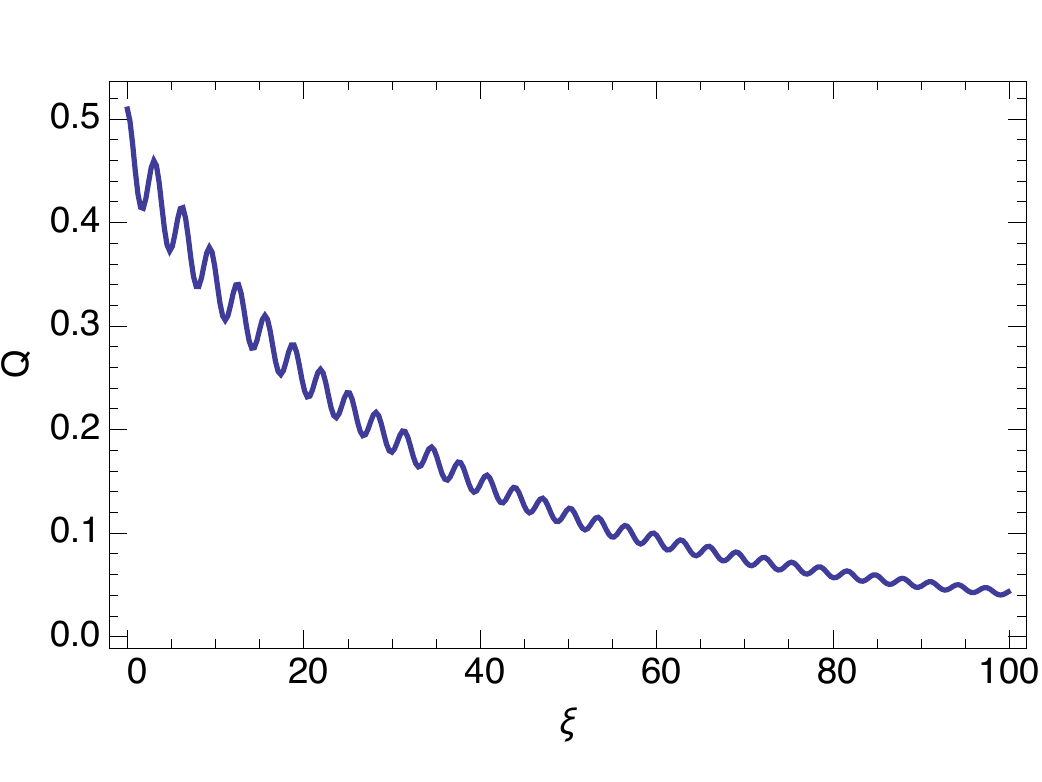}
    \caption{The plot of the function $Q$ versus $\xi = s/\beta$ for a beam with $\tau = 5\times 10^{-4}$ for $K = 1$.}
    \label{fig:5}
    \end{figure}

The large-scale correlations  with small values of $K$ decay on a relatively large distance. This is illustrated by Fig.~\ref{fig:5} which shows the function $Q$ for $K = 1$.
We see that it takes many betatron periods for $Q$ to decay to small values in this case.    
    
\begin{figure}[htb!]
    \centering
    \includegraphics[width=0.5\textwidth, trim=0mm 0mm 0mm 0mm, clip]{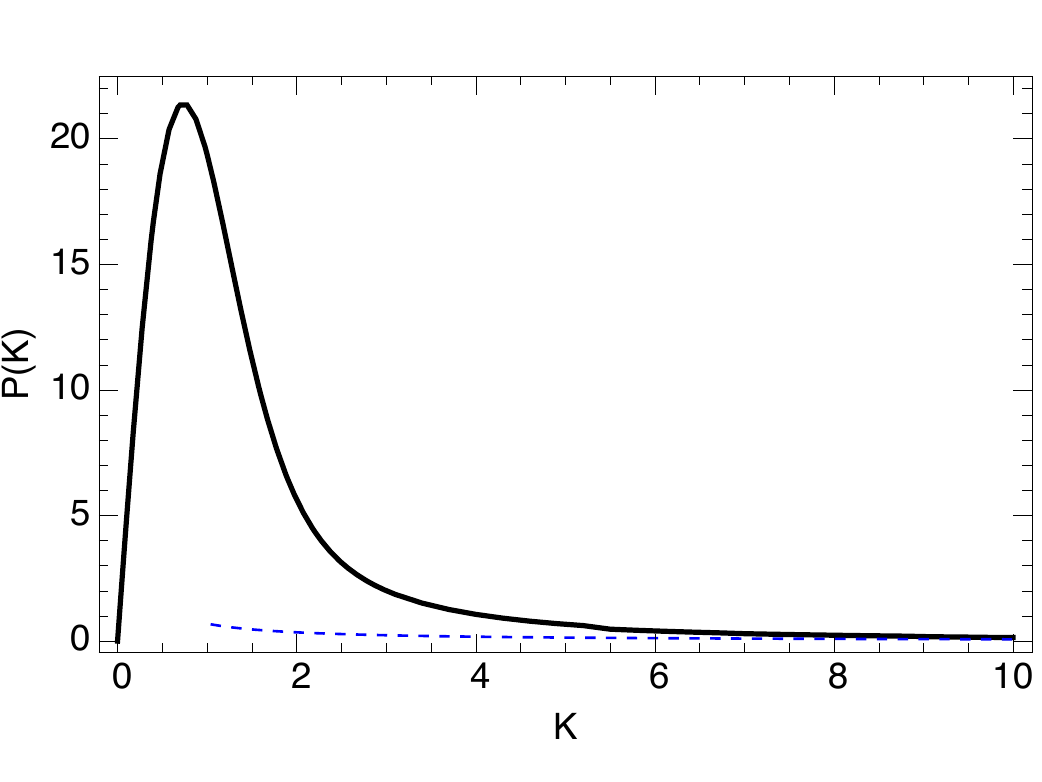}
    \caption{The plot of the integral $P(K)$ (black line). At large values of $K$ this integral decays as $\pi^{3/2} /8K$ shown by the blue dashed line.}
    \label{fig:6}
    \end{figure}

Finally, Fig.~\ref{fig:6} shows numerically calculated function $P(K)$  for our case ($\sigma_\eta = 10^{-4}$). Theoretical analysis of Eqs.~\eqref{eq:48} and~\eqref{eq:50} reveals that at large values of $K$, $P(K)$ decays as $\pi^{3/2} /8K$, which is in agreement with numerical calculations shown in Fig.~\ref{fig:6}.    
According to Eq.~\eqref{eq:48}, to find $J$, we need to integrate $P(K)$ over $K$ from zero to infinity. This can be done by integrating the numerically calculated $P(K)$ from zero to some large value $K_0$ and then using the asymptotic formula $P(K) \approx \pi^{3/2} /8K$ for $K > K_0$. Of course, such an  integral logarithmically diverges at infinity and, as it was done in Section~\ref{sec:5}, we need to truncate the integration interval by some value $K_\mathrm{max} \gg 1$. Carrying our this calculation, we obtain the following result:
	\begin{align}\label{eq:51}
	J
	=
	33.3
	+
	\frac{\pi^{3/2}}{8}
	\log(K_\mathrm{max} \sqrt{\epsilon \beta})
	.
	\end{align}
The second term in this expression substituted to Eq.~\eqref{eq:46} gives the energy diffusion coefficient that  (apart of somewhat different definition of the log term) coincides with Eq.~\eqref{eq:46} and Eq.~\eqref{eq:5}. The first term in Eq.~\eqref{eq:51} adds a noticeable correction to the Coulomb logarithm due to the long-range correlations of the fluctuating fields in the beam.

%
\section{Summary and discussion}\label{sec:8}
%

In this work, we derived the energy diffusion coefficient due to the intra-beam scattering in a relativistic beam. In contrast to the conventional approach, we used the framework of the fluctuations of the electric field in the beam caused by random motion of  beam particles. We also included the effect of betatron oscillations on the collisions in our analysis. 

In Section~\ref{sec:4}, we derived a general expression for the energy diffusion coefficient $\bar{\cal D}$ averaged over the distribution function of the beam. Quite surprisingly, an attempt to calculate its value for a beam with zero energy spread resulted in a diverging integral. In Section~\ref{sec:5}, we have shown that the situation can be rectified if we limit the integration domain by wavenumbers of the fluctuating fields that are much smaller than the transverse beam size. In this case, we obtained the result that coincides with the conventional formula~\eqref{eq:5} with a logarithmic factor that mimics the Coulomb logarithm. Such truncation of the integration range means that Eq.~\eqref{eq:5} ignores the long-range fluctuations that exist in the beam and amplifies  the diffusion caused by the short-range fluctuations.
 
A numerical example presented in Section~\ref{sec:6} clearly demonstrated the presence of such correlations in the field correlator $Q$. For a finite energy spread shown in Fig.~\ref{fig:2}, these correlations decay with distance, however, in the limit $\sigma_\eta = 0$ they persist without damping as shown in Fig.~\ref{fig:4}. In the latter case, these repeated spikes in the correlations lead to a divergent integral in the expression for $\bar{\cal D}$. For $\sigma_\eta \ne 0$, the extended correlations manifest itself in the increase of the effective Coulomb logarithm as demonstrated by Eq.~\eqref{eq:51}.

\begin{figure}[htb!]
    \centering
    \includegraphics[width=0.7\textwidth, trim=0mm 0mm 0mm 0mm, clip]{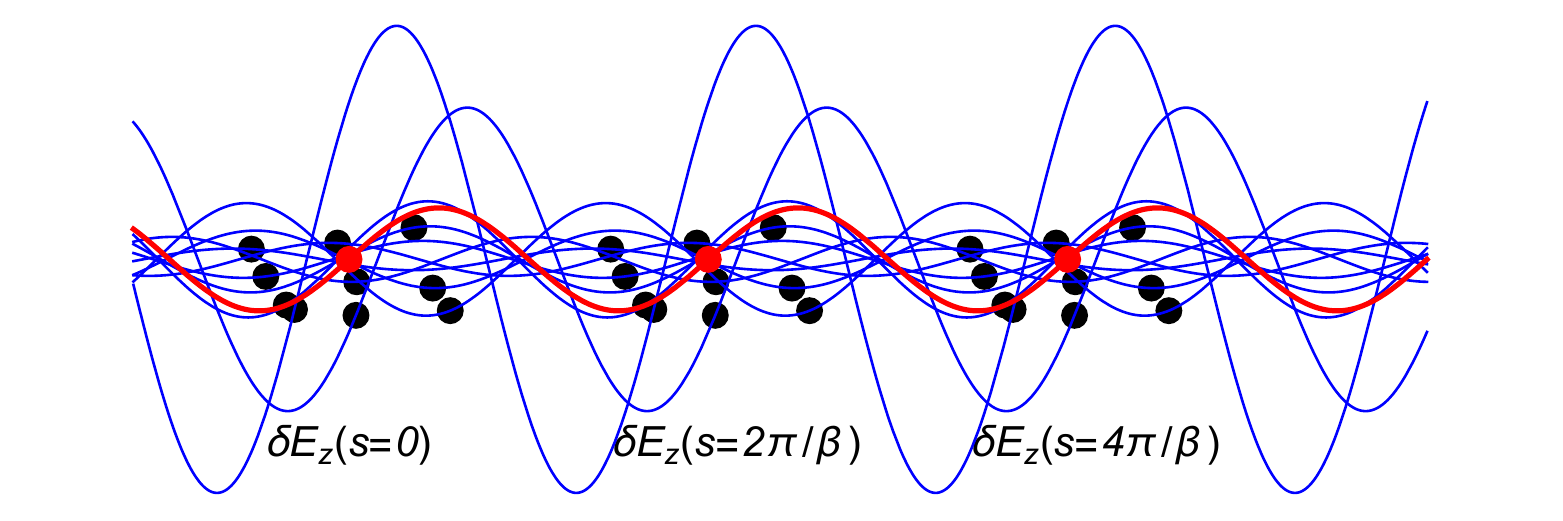}
    \caption{The plot of the positions of the test particle (red dot) and the source particles (black dots) executing betatron oscillacion. The red line is the trajectory of the test particle and the blue lines are the trajectories of the source particles. The positions of all particles repeat every betatron period.}
    \label{fig:7}
    \end{figure}

In our calculations, we have assumed a smooth focusing with the equal betatron oscillation frequencies in $x$ and $y$ directions. In this situation, the microscopic positions of the particles arrange themselves into the same pattern after each betatron period (and the positions are mirror symmetric after half a period). This is illustrated in Fig.~\ref{fig:7}.
In the language of binary collisions, such patterns mean that a collision between two particles will be repeated every betatron period, and the result of such collisions adds up in a regular fashion instead of being accumulated as random encounters. This also means that in this case the diffusion-like scaling in the growth of the energy spread in the beam due to IBS, $\Delta\sigma_\eta \propto s$, is not valid; it will be replaced by $\Delta\sigma_\eta \propto s^2$.

The persistence of the long-range correlations in the electric field  of the beam has been demonstrated in this paper for very special model of the beam transport. In reality, in addition to the beam energy spread, there will be other mechanisms that will be suppressing these correlations, e.g., the lattice chromaticity and nonlinearity, quantum diffusion in the beam, etc. Their influence of these factors on the field correlations requires further studies..

Finally, we note that even though we considered only the energy diffusion in this paper, other elements of the collision integral will be affected by the long-scale correlations. In particular, on can probably see their effect in the emittance growth of the beam that is crucial in the modern light sources.

%
\section{Acknowledgements}\label{sec:8}
%

I would like to thank Z. Huang, R. Robles and S. Kladov for useful discussions. 

This  work was supported  by  the  Director,  Office  of  Science,  Office  of  Basic  Energy  Sciences,  of  the  U.S.  Department  of  Energy  under  Contracts  No. DE-AC02-76SF00515.

This work

\appendix

%
\section {Calculation of the longitudinal field in the beam}\label{app:A}
%

We consider the beam moving with velocity $v_0 = \beta_0 c$ along the $z$ axes. Let the charge density perturbation of the beam be $e \delta n(x,y,z)$, then one can obtain the following equation for the longitudinal component of the electric field from Maxwell's equations,
    \begin{align}\label{eq:A.1}
    \nabla_\perp \delta E_z
    +
    \frac{1}{\gamma^2}
    \p_{zz} \delta E_z
    =
    \frac{4\pi e}{\gamma^2}
    \p_{z}\delta n
    , 
    \end{align}
where $z = s-v_0t$, $\nabla_\perp$ is the transverse gradient operator, and $\p_{z}$ and $\p_{zz}$ denote the first and the second partial derivatives along $z$. Making the 3D Fourier transformation,
    \begin{align}\label{eq:A.2}
    \delta \hat n_{\vec k}
    &=
    \intinf
    dx\,dy\,dz\,
    e^{-i(k_x x+k_y y+k_zz)}
    \delta n(x,y,z)
    ,
    \nonumber\\
    \delta \hat E_{z,\vec k}
    &=
    \intinf
    dx\,dy\,dz\,
    e^{-i(k_x x+k_y y+k_zz)}
    \delta E_z(x,y,z)
    ,
    \end{align}
we obtain
    \begin{align}\label{eq:A.3}
    \delta \hat E_{z,\vec k}
    =
    \frac{4\pi e i k_z}{\gamma^2(k_x^2+k_y^2)+k_z^2}
    \delta \hat n_{\vec k}
    .
    \end{align}

The energy modulation in Fourier representation (per unit length of path) is
    \begin{align}\label{eq:A.4}
    \delta \hat \eta_{\vec k}
    =
    -e\hat \delta E_{z,\vec k}
    \frac{1}{m_e\gamma c^2}
    =
    -
    \frac{4\pi e^2 i k_z}{\gamma^2(k_x^2+k_y^2)+k_z^2}
    \frac{1}{m_e\gamma c^2}
    \delta \hat n_{\vec k}
    .
    \end{align}

%
\section{Density fluctuations in the beam}\label{app:B}
%

To calculate the density fluctuations, we have to start from the following microscopic (Klimontovich) distribution function describing a collection of non-interacting particles that execute betatron oscillations,
    \begin{align}\label{eq:B.1}
    & f_\mathrm{M}(J_x,\phi_x,J_y,\phi_y,\eta,\zeta,s)
    =
    \frac{1}{N}
    \sum_{n=1}^N
    \delta (J_x-J_{x,n})
    \delta (\phi_x - \psi_x(s) - \phi_{x,n})
    \\\nonumber&
    \delta (J_y-J_{y,n})
    \delta (\phi_y - \psi_y(s) - \phi_{y,n})
    \delta(\eta-\eta_n)\delta(\zeta - \zeta_{n})
    ,
    \end{align}
where $N \gg 1$ is the number of particles in the beam. This distribution function is normalized so that $\int dJ_x\, dJ_y\,d\phi_x\,d\phi_y\,d\eta\,d\zeta\,f_\mathrm{M} = 1$. 

The 6N parameters in Eq.~\eqref{eq:B.1}, $J_{x,n}, \phi_{x,n}, J_{y,n},\phi_{y,n},\zeta_{n},\eta_{n}$, are random variables that are distributed according to the averaged distribution function $f_0$. Specifically, the phases $\phi_{x,n}$ and $\phi_{y,n}$ are uniformly distributed in the interval $[0, 2\pi]$, the longitudinal coordinate $\zeta$ is uniformly distributed over $[0, L]$, and the variables $J_{x,n}, J_{y,n}$ and $\eta_{n}$ are distributed according the distribution functions $f_{0x}$, $f_{0y}$ and $f_{0z}$, respectively.  We will denote by the angular brackets $\langle \ldots \rangle$ averaging over these variabless:
    \begin{align}\label{eq:B.2}
    \langle
    \ldots
    \rangle
    &=
    \prod_{n=1}^N
    \int_0^\infty 
    f_{0x}(J_{x,n})\,
    dJ_{x,n}
    \int_0^{2\pi}
    d\phi_{x,n}
    \int_0^\infty dJ_{y,n}
    \, f_{0y}(J_{y,n})
    \int_0^{2\pi}
    d\phi_{y,n}
    \nonumber\\
    &
    \times
    \int_0^L d \zeta_{n}
    \intinf
    d\eta_{n}
    f_{0z}(\eta_{n})
    \ldots
    .
    \end{align}
Applying this averaging to $f_\mathrm{M}$ gives the averaged distribution function, $\langle f_\mathrm{M} \rangle = f_0(J)$. In our calculations, we will use Eqs.~\eqref{eq:24},~\eqref{eq:25} for $f_0$. Note that the longitudinal distribution function is normalized by
\begin{align}\label{eq:B.3}
\int_0^L 
d\zeta
\intinf
d\eta
f_{0z}(\eta)
=
L
\intinf
d\eta
f_{0z}(\eta)
=
1
,
\end{align}
where $L$ is the length of the bunch (in this analysis we assume a uniform longitudinal distribution of particles).

Subtracting from $f_\mathrm{M}$ its average value gives the perturbation of the distirbution function $\delta f_\mathrm{M}$,
    \begin{align}\label{eq:B.4}
    \delta f_\mathrm{M} = f_\mathrm{M} - f_0,
    \end{align}
where $\delta f_\mathrm{M}$ is small, $|\delta f_\mathrm{M}| \ll f_0$. The function  $\delta f_\mathrm{M}$ is responsible for the density fluctuations in the beam. Being a function of random variables $J_{x,n}, \phi_{x,n}, J_{y,n},\phi_{y,n},\zeta_{n},\eta_{n}$, it itself is  a random variable.  Integrating it over the whole phase space with the proper delta-functions that establishes a relation beteen particle's  coordinates  $x, y, z$ and the action-angle variables (see Eqs.~\eqref{eq:6} and Eq.~\eqref{eq:9}), gives the density perturbation $\delta n(x,y,z,s)$, 
    \begin{align}\label{eq:B.5}
    \delta n(x,y,z,s)
    &=
    N
    \int_0^\infty dJ_{x}
    \int_0^{2\pi}
    d\phi_{x}
    \int_0^\infty dJ_{y}
    \int_0^{2\pi}
    d\phi_{y}
    \int_0^L 
    d\zeta
    \intinf
    d\eta\,
    \delta f_\mathrm{M}(J_x,\phi_x,J_y,\phi_y,\eta,\zeta,s)
    \nonumber\\&\times
    \delta(x-\sqrt{2\beta_x(s)J_x}\cos(\psi_x(s)+\phi_x))
    \delta(y-\sqrt{2\beta_y(s)J_y}\cos(\psi_y(s)+\phi_y))
    \nonumber\\&\times
    \delta\left(z-\zeta-\frac{s\eta}{\gamma^2}\right)    
    .
    \end{align}
As is $\delta f_\mathrm{M}$, $\delta n$ is a random function, and we are interested in calculation the  averaged product $\langle\delta n(x,y,z,s)\delta n(x',y',z',s')\rangle$. A straightforward although lengthy calculation that uses Eqs.~\eqref{eq:B.1}--\eqref{eq:B.4} gives the following result:
    \begin{align}\label{eq:B.6}
    &\langle\delta n(x,y,z,s)\delta n(x',y',z',s')\rangle
    = N\times
    \nonumber\\&
    \int_0^\infty dJ_{x}
    \int_0^{2\pi}
    d\phi_{x}
    f_{0x}(J_{x})
    \delta(x-\sqrt{2\beta_x(s)J_{x}}\cos[\psi_x(s) + \phi_{x}])
    \delta(x'-\sqrt{2\beta_x(s')J_{x}}\cos[\psi_x(s') + \phi_{x}])
    \nonumber\\
    &
    \int_0^\infty dJ_{y}
    \int_0^{2\pi}
    d\phi_{y}
    f_{0y}(J_{y})
    \delta(y-\sqrt{2\beta_y(s)J_{y}}\cos[\psi_y(s) + \phi_{y}])
    \delta(y'-\sqrt{2\beta_y(s')J_{y}}\cos[\psi_y(s') + \phi_{y}])
    \nonumber\\&\times
    \intinf
    d\eta
    f_{0z}(\eta)    
    \delta\left(z-z'-\frac{(s-s')\eta}{\gamma^2}\right)    
    .
    \end{align}

The Fourier transform of the density fluctuations is defined by Eq.~\eqref{eq:21}. For the correlator of the Fourier components, $\langle    \delta \hat n_{\vec k}(s)    \delta \hat n_{\vec k'}(s')    \rangle$, through which we have expressed the correlator of the electric field in Eq.~\eqref{eq:22}, we have
    \begin{align}\label{eq:B.7}
    &\langle \delta \hat n_{\vec k}(s)    \delta \hat n_{\vec k'}(s')    \rangle
    =    
    \int
    dx\,dy\,dz\,dx'\,dy'\,dz'\,
    e^{i{\vec k\cdot \vec r} + i{\vec k'\cdot \vec r'}}
    \langle\delta n(x,y,z,s)\delta n(x',y',z',s')\rangle
    .
    \end{align}
    Substituting Eq.~\eqref{eq:B.6} into this equation and taking the integrals gives the final result:
    \begin{align}\label{eq:B.8}
    &\langle
    \delta \hat n_{\vec k}(s)
    \delta \hat n_{\vec k'}(s')
    \rangle
    =
    2\pi\nu
    \delta(k_z+k'_z)
    \exp\left[
    -\frac{1}{2}\sigma_\eta^2\left(
    k_z\frac{(s-s')}{\gamma^2}
    \right)^2
    \right]    
    \nonumber\\&\times
    \exp
    \left[
    -
    \frac{1}{2}
    \epsilon_x
    \left(
    k_x^2\beta_x(s)
    +
    k_x'^2\beta_x(s')
    +
    2k_xk_x'\sqrt{\beta_x(s)\beta_x(s')}\cos(\psi_x(s) - \psi_x(s'))
    \right)
    \right]
    \nonumber\\
    &\times
    \exp
    \left[
    -
    \frac{1}{2}
    \epsilon_y
    \left(
    k_y^2\beta_y(s)
    +
    k_y'^2\beta_y(s')
    +
    2k_yk_y'\sqrt{\beta_y(s)\beta_y(s')}\cos(\psi_y(s) - \psi_y(s'))
    \right)
    \right]
    \end{align}
where $\nu = N/L$ is the number of particles per unit length.
For the correlation at one location, $s=s'$, we obtain
    \begin{align}\label{eq:B.9}
    \langle
    \delta \hat n_{\vec k}
    \delta \hat n_{\vec k'}
    \rangle
    &=
    2\pi\nu
    \delta(k_z+k'_z)
    \exp\left[
    -\frac{1}{2}\sigma_\eta^2\left(
    k_z\frac{(s-s')}{\gamma^2}
    \right)^2
    \right]
    \nonumber\\&\times
    \exp
    \left[
    -
    \frac{1}{2}
    \sigma_{x}^2
    \left(
    k_x
    +
    k_x'
    \right)^2
    \right]
    \exp
    \left[
    -
    \frac{1}{2}
    \sigma_{y}^2
    \left(
    k_y
    +
    k_y'
    \right)^2
    \right]
    .
    \end{align}

%
%

\bibliography{\string~/gsfiles/Bibliography/master%
}

\end{document}